\def\puncspace{\ifmmode\,\else{\ifcat.\C{\if.\C\else\if,\C\else\if?\C\else%
\if:\C\else\if;\C\else\if-\C\else\if)\C\else\if/\C\else\if]\C\else\if'\C%
\else\space\fi\fi\fi\fi\fi\fi\fi\fi\fi\fi}%
\else\if\empty\C\else\if\space\C\else\space\fi\fi\fi}\fi}
\def\SP{\let\\=\empty\futurelet\C\puncspace}

\def\h1{$h^{-1}$\SP}

\def\etal{{\it et al.\/}\ }

\def\eg{{\it e.g.\/}\rm,\ }
\def\lsim{~\rlap{$<$}{\lower 1.0ex\hbox{$\sim$}}}
\def\gsim{~\rlap{$>$}{\lower 1.0ex\hbox{$\sim$}}}
\def\void#1{{}}

\documentclass{aa}
\usepackage{graphics}
\begin{document}

   \thesaurus{03 % A&A Section 3: Extragalactic Astronomy
               (11.05.1;
                11.05.2;
                11.06.1;
                12.03.3)}     

%              (03.11.1)}  % Cosmogony,
%               16.06.1;  % Planets and satellites: general,
%              19.06.1;  % Solar system: general,
%               19.37.1;  % Stars: formation of,
%               19.53.1;  % Stars: oscillations of,
%               19.63.1)} % Stars: structure of.
%
   \title{EROs in the EIS Fields}

   \subtitle{I. The AXAF (Chandra) Deep Field
\thanks{Based on observations collected at the European Southern
Observatory, La Silla, Chile; ESO programs N$^o$ 61.A-9005 and 162.O-0917}
}

\author {Marco Scodeggio\thanks{current address: Istituto di Fisica
Cosmica ``G. Occhialini'', via Bassini 15, I-20133, Milano, Italy}
\and David R. Silva} 

%   \author{EIS Team
%          \inst{1}
%          \and
%          C. Ptolemy\inst{2}\fnmsep\thanks{Just to show the usage
%          of the elements in the author field}
%          }

   \offprints{Marco Scodeggio}

\institute{
European Southern Observatory, Karl-Schwarzschild-Str. 2,
D--85748 Garching b. M\"unchen, Germany\\
email: mscodegg@eso.org (marcos@ifctr.mi.cnr.it), dsilva@eso.org
}

%   \institute{European Southern Observatory, Karl-Schwazschild-Str. 2
%    Garching bei M\"unchen, Germany\\
%              email: 
%         \and
%             University of Alexandria, Department of Geography\\
%             email: c.ptolemy@hipparch.uheaven.space
%             \thanks{The university of heaven temporarily does not
%                     accept e-mails}
%             }

   \date{Received ; accepted }

   \maketitle

\begin{abstract}

The publicly available EIS-DEEP optical-NIR data for the AXAF
(Chandra) Deep Field have been used to construct samples of Extremely
Red Objects (EROs) using various single-band and multi-band color
criteria.  In this work we define as EROs objects with colors
consistent with passively evolving elliptical galaxies at z $\geq$ 1.
The EROs surface densities we derive are intermediate between previous
published values, emphasizing again the need for larger survey areas
to constrain the effects of possible large-scale structure.  Although
various single-color selected samples can be derived, the EROs sample
selected using $R-Ks > 5, I-Ks > 4, J-Ks > 1.8$ jointly is likely to
contain the highest fraction of passively evolving luminous field
elliptical galaxies at z$\geq$1, or conversely, the lowest fraction of
lower redshift interlopers.  The surface density of this multi-band
selected EROs sample is consistent with the conclusion that little or
no field elliptical volume density evolution has occurred in the
redshift range 0 $>$ z $>$ 1.5.  However, extensive spectroscopic
followup is necessary to confirm this conclusion.

\end{abstract}

%________________________________________________________________

\section{Introduction}
\label{sec:intro}

The formation history of field elliptical galaxies (i.e. elliptical
galaxies located outside of massive galaxy clusters) remains very
uncertain.  Although most stellar population indicators point towards
star formation at z $>$ 2, it remains unclear when and how these
stars were assembled into their current dynamical structures.

Two formation scenarios with distinct observational predictions
continue to be debated.  Each one allows for star formation to take
place at high redshift, i.e.  for the present epoch stellar
populations of field ellipticals to be mostly old and metal-rich.
However, the first scenario, based on the {\it rapid dissipative
collapse} model, postulates that the stars were formed and assembled
into elliptical galaxies in one rapid collapse event at high (z $>$ 2)
redshift (Eggen, Lynden-Bell, \& Sandage 1962; Larson 1974; Tinsley
\& Gunn 1976).  In this ``top-down'' scenario field ellipticals have 
evolved quiescently since their formation, and therefore the number of 
massive/luminous field elliptical per co-moving volume should be 
constant at all redshifts between now and their formation epoch.  On 
the contrary, the opposing {\it hierarchical merger} scenario assumes 
that ellipticals were assembled from previously formed sub-units, and 
that this process has continued until the current epoch (see, e.g.  
Toomre 1977; White \& Rees 1978; Blumenthal et al.  1984).  In this 
``bottom-up'' scenario, the number of massive/luminous elliptical 
field galaxies per co-moving volume increases over time.

These seemingly distinct and easily tested observational predictions
have in practice been difficult to resolve at relatively low redshift.
Based on the Canada-France Redshift Survey, Lilly et al.  (1995)
concluded that there has been no significant number evolution for
bright, early-type galaxies since z $=$ 1. This result was questioned
by Kauffmann, Charlot, \& White (1996), but it has been confirmed by
Schade \etal (1999), who took advantage in their study also of the
availability of detailed morphological information derived from HST
images.  Comparing the co-moving number density of field elliptical
galaxy at z $>$ 1 to the current epoch number density would be more
fruitful because the longer time baseline creates a larger (and
therefore easier to detect) number density difference in the
hierarchical model.  Under the rapid dissipative collapse model, of
course, the co-moving number density would remain constant.

The availability of large format near-IR arrays has created the
opportunity to search for the luminous field ellipticals at high
redshift needed to carry-out this test.  These searches are based on a
very simple concept: once the 4000 \AA\ break is redshifted beyond the
I-band, the observed optical-IR colors of a quiescently evolving
elliptical become very red (R--K $>$ 5; I--K $>$ 4).  In terms of
brightness, a typical current epoch field elliptical redshifted to z
$>$ 1 will have an observed K of $\approx$ 19 -- 21 (Moustakas et al.
1997).  Objects with these extreme properties (faint K-band
magnitudes, extremely red optical-NIR colors) are now commonly known
as Extremely Red Objects (EROs).

Of course, EROs imaging surveys only measure EROs surface density.
Determining the field elliptical galaxies volume density ultimately
requires a combination of morphological and spectroscopic information
to eliminate Galactic and extragalactic interlopers and to determine
the redshifts of high redshift field elliptical galaxy candidates.
Although some EROs have been found to have spectral properties
consistent with young (3 - 4 Gyr) elliptical galaxies (\eg Cowie \etal
1996; Dunlop \etal 1996; Spinrad \etal 1997; Cohen \etal 1999),
spectra and/or exact redshifts are available for only a very small
fraction (10 -- 15\%) of detected EROs.  Thus, it is possible that
some EROs are in fact dust-enshrouded galaxies at lower redshift (as
discussed, e.g., by Graham \& Dey 1996; Cimatti \etal 1997) or
foreground lower main sequence and brown dwarf stars, although the
latter possibility is extremely unlikely given the apparent surface
density of such Galactic objects (Moustakas et et.  1997; Cohen \etal
1999).  Despite such contamination problems, observed EROs number
surface density should be able to place a stringent upper limit on
luminous field elliptical galaxy number volume density at z $>$ 1.

\begin{table*}
    \caption{Published Deep EROs Survey Results}
    \label{tab:surveys}
\begin{tabular}{l c c c c c c}
\hline \hline
\hfil Reference \hfil&
Color &
Area & 
$K_{\rm lim}$ & 
N(EROs) &
N(EROs)/deg$^2$ \\
\hfil (1) \hfil & (2) & (3) & (4) & (5) & (6) \\
\hline
Djorgovski et al. 1995  & i--K & 3  & 20 &  1 & 1200 $\pm$ 1200  \\
Cowie et al. 1996       & I--K & 26 & 20 & 19 & 2611 $\pm$  600  \\
Moustakas et al. 1997   & I--K & 2  & 22 & 11 & 19800$^{\mathrm{a}}$
$\pm$ 6000 \\
Moustakas et al. 1997   & I--K & 2  & 20 &  3 & 5400 $\pm$ 3100  \\
Barger et al. 1999      & I--K & 62 & 20 & 16 &  929 $\pm$  230  \\
Cohen et al. 1999       & R--K & 15 & 20 & 19 & 4560 $\pm$ 1050  \\
Thompson et al. 1999    & R--K &154 & 20 &289 & 6750 $\pm$  400  \\
This work               & I--K & 23 & 20 & 20 & 3130 $\pm$  700  \\
This work               & R--K & 43 & 20 & 22 & 1842 $\pm$  390  \\
This work &RIJK$^{\mathrm{b}}$ & 23 & 20 & 10 & 1565 $\pm$  495  \\
\hline \hline
\end{tabular}

{\bf Columns:} (1) Reference; (2) Color Criteria (R--K $>$ 5 or I--K
$>$ 4); (3) Area Surveyed (square arcminutes); (4) K-band EROs
Limiting Magnitude; (5) Number of EROs Found; (6) Implied Number of
EROs per square degree, with associated Poissonian uncertainty.
\begin{list}{}{}
\item[$^{\mathrm{a}}$] Note the different limiting magnitude
associated with this EROs density.
\item[$^{\mathrm{b}}$] This is a multi-color selection criterion,
discussed in Sect. 4.
\end{list}

\end{table*}

Unfortunately, published results from such surveys do not present a 
consistent picture of EROs surface density, let alone inferred volume 
density.  As Table~\ref{tab:surveys} illustrates, measured EROs surface 
densities vary by factors of 2 -- 6.  Clearly, a consensus result has 
not been reached yet.  However, given that most of these surveys cover 
very small areas, it is not impossible that the observed surface density 
variations are caused by large-scale structure at high redshift.  
Additional samples, and preferable over larger areas, are still 
needed.

In this first in a series of three papers, we present an analysis of
the optical-NIR images obtained of the AXAF (Chandra) Deep Field as
part of the ESO Imaging Survey (EIS; Rengelink et al. 1999).  These
data cover an area of approximately 43 square arcmin, comparable in
spatial area to the largest EROs surveys already published.  The EIS
data have been used to construct EROs samples using a variety of
magnitude and color criteria.  In this paper, we review the EIS
dataset (Sect.~\ref{sec:data}) and then present a number of possible
EROs samples using different selection criteria
(Sect.~\ref{sec:samples}).  The sample most likely to represent field
elliptical galaxies at z $>$ 1 is then discussed in
Sect.~\ref{sec:ellipticals}.  Our results are summarized in
Sect.~\ref{sec:summary}.

In the next two papers, we will present an analogous analysis of the 
public EIS-DEEP HDF-S dataset (da Costa et al.  1999) and of a private 
near-IR dataset obtained as part of two independent EIS Cluster 
candidate followup surveys.  The total area surveyed by this joint 
study will be $\sim$ 400  square arcmin, i.e.  larger than any other 
published EROs imaging survey at a comparable K-band limiting 
magnitude.

\section{The data}
\label{sec:data}

This work is based on the publicly available data obtained as part of
the ESO Imaging Survey (EIS) on the so-called AXAF (Chandra) Deep
Field, at $\alpha = 03^h 32^m30^s$ and $\delta = -27^{\circ}48'30''$.
Deep multi-band optical and infrared observations in this field were
obtained in the period August-November 1998, using the ESO 3.5m New
Technology Telescope (see Rengelink \etal 1999 for details).  Optical
observations in the $UBVRI$ bands were carried out with the SUSI2
camera (D'Odorico \etal 1998), equipped with two $4k \times 2k$ EEV
CCDs, covering a field of view of $5.46
\times 5.46$ arcmin with a pixel scale of 0.16 arcsec/pixel (after $2 
\times 2$ binning).  Infrared observations in the $J$ and $Ks$ bands 
were carried out using the SOFI infrared camera and spectrograph 
(Moorwood, Cuby \& Lidman 1998), equipped with a Rockwell $1k \times 
1k$ detector, covering a field of view of $4.9 \times 4.9$ arcmin with 
a pixel scale of 0.29 arcsec/pixel.

The infrared observations cover 4 SOFI fields, for a total area of 
approximately 83 square arcmin.  Optical observations at the moment 
cover only the two northern SOFI fields, for a total area of 
approximately 56 square arcmin, except in the I band, where only a 
smaller 30 square arcmin area has been covered.  Total exposure times 
are 5500 seconds for the $V$- and $R$-band exposures, 12600 seconds 
for the $I$-band exposure, and 10800 seconds for the $J$- and 
$Ks$-band exposures.  Details about the data reduction procedures are 
given by Rengelink \etal (1999), and da Costa \etal (1999), and only a 
brief summary is presented here.  Single dithered optical exposures 
were coadded using the drizzle procedure (Fruchter \& Hook 1998) 
implemented within the EIS pipeline.  Infrared jittered images were 
combined using the {\it jitter} program within the Eclipse 
data-reduction package (Devillard 1998).  The final single-band 
coadded images were astrometrically calibrated using the USNO-A V1.0 
catalog as a reference.  The photometric calibration was based on 
observations of Landolt (1992) standard stars for the optical data, 
and of HST standards from the list of Persson \etal (1998) for the 
infrared data.  Zero-point uncertainties are of $\pm0.03$ mag in 
$V$-band, $\pm 0.04$ mag in $R$-band, and of $\pm 0.05$ mag in $I$-, 
$J$-, and $Ks$-band.

A multi-color object catalog was created based on the method of the
chi-squared image, described by Szalay, Connoly \& Szokoly (1998).
All single-band coadded images were convolved with a Gaussian kernel
to devise a set of images with homogeneous PSFs (i.e.  equivalent to
the single-band image with worse seeing, the I-band image), normalized
by their respective {\it rms} noise values, and then quadratically
combined to obtained a chi-squared image.  Object detection on that
image was carried out using the SExtractor software (Bertin \& Arnouts
1996).  Magnitude measurements based on the detection parameters
derived from the chi-squared image were performed on the Gaussian
convolved single-band coadded images.\footnote{Our final catalog
differs somewhat, especially at the faintest magnitude, from the AXAF
Deep Field catalogs released by the EIS Team in December 1998 on their
Web site.  When constructing the December 1998 catalogs, the coadded
single-band images were not convolved by a Gaussian before the
chi-square image was constructed and object magnitudes were measured
on the non-convolved single-band images.} Magnitudes were then
measured within a 4 arcsec diameter aperture.  From a comparison with
deep number counts in the literature, it is estimated that the 90\%
completeness limit of the multicolor catalog in the different bands is
approximately $V = 26.2$, $R = 26.0$, $I = 25.6$, $J = 23.6$, $Ks =
21.6$ (see da Costa \etal 1999, and Rengelink \etal 1999). Star/galaxy
separation was based on the SExtractor stellarity index, considering
as galaxies object with index $\leq 0.85$, when measured in the
$Ks$-band images. It is important to remark that this classification
criterion is reliable only for galaxies at least 1--1.5 mag brighter
than the detection limit. Stellar contamination is however expected to
be very limited at this galactic latitude ($b \simeq -55$), and our
selection of very red objects is expected to reduce it even further
(see for example da Costa \etal 1999).

In this work, we restrict ourselves to the area covered by both the
optical and infrared observations, and further limit our analysis to the
area where the sensitivity of the single-band coadded images is most
uniform, rejecting those parts of the surveyed area where the
sensitivity is below 75\% of the peak in any of the single-band coadded
images.  In this way, the area we consider covered by the $RJKs$ 
observations is $\sim$ 43 square arcmin, while that covered by the 
$RIJKs$ observations is $\sim$ 23.5 square arcmin.  Within these areas 
we derive an object catalog by selecting all objects measured above 
the 2$\sigma$ limit in the $Ks$-band, which in practice limited us to 
objects brighter than $Ks =$ 21.0.  At that $Ks$ limiting magnitude, 
the equivalent limiting magnitudes in our other bands were $J = 22.8$, 
$I = 25.0$, and $R = 26.0$ with magnitude uncertainties of 
approximately 0.1 mag in $R$, and 0.15 mag in $I$, $J$, and $Ks$.

\section{Single-color selected EROs samples} 
\label{sec:samples} 

\begin{figure}
\resizebox{\columnwidth}{!}{\includegraphics{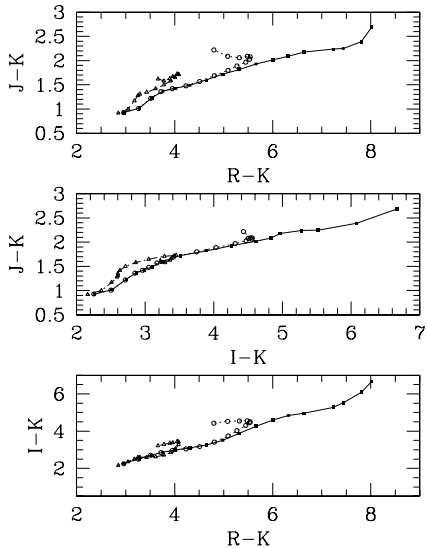}} 
\caption{Color--color tracks for galaxies with different star formation
histories. The observed colors as a function of galaxy redshift, in
the interval $0 < z < 2$, are plotted for an elliptical galaxy (solid
line, filled squares), an early-type spiral (dotted line, open
circles), and a late-type spiral (dashed line, open triangles). The
dots identify redshift intervals $\Delta z = 0.1$, starting from the
bottom-left in each panel. The colors are computed using population
synthesis models from Bruzual \& Charlot (1993), assuming an H$_0$=75
km s$^{-1}$Mpc$^{-1}$, $\Omega_0=0.3$ cosmological model, Salpeter IMF,
and solar metallicity for all galaxies.}
\label{fig:color_color}
\end{figure}

As our main objective is to study the evolution of the volume density
of $z>1$ field elliptical galaxies, we define EROs to be
color-selected objects that are redder than a z $\simeq$ 1 passively
evolving elliptical galaxy.  While this definition has been already
used by other authors (\eg Barger \etal 1999, Cohen \etal 1999), it is
not unanimously adopted in the literature, as other studies have
defined EROs to be objects that are redder than a passively evolving
elliptical galaxy at any redshift (\eg Andreani \etal 1999, Thompson
\etal 1999). Since most previously published studies have relied on 
a single color criterion to select EROs, we decided to select from our
dataset a number of different single color samples, based on the
criteria $R-K \geq 5.0$, $I-K \geq 4.0$, or $J-K \geq 1.8$. We make no
use of the available V-band data since it is too shallow for our
requirements (the color criterion would be $V-K \geq 7.0$, which would
require completeness down to V = 28). These criteria were designed to
include in the samples all the elliptical galaxies at $z \geq 1$ that
would have properties comparable to those of their present-day
counterparts, taking into account the passive evolution of their
stellar population (assumed to have formed at $z > 2$). At the same
time, these choices allow comparisons with previously published EROs
surveys to be readily carried out, although the effective redshift
cutoff for passively evolving early-type galaxies might be slightly
different for the different colors, as the color-color tracks plotted
in Figure~\ref{fig:color_color} show. The effect however is very
small, and does not significantly affect the analysis carried out in
this paper.

Obviously a single color measurement is not enough to define, even
approximately, a galaxy's spectral energy distribution (SED).
Therefore every EROs sample that is defined on the basis of a
single-color criterion is likely to contain both $z\geq 1$ early-type
galaxies and a mixture of other objects, including dusty star-forming
galaxies, and low-mass stars. Moreover, as illustrated in Figure
~\ref{fig:color_color}, samples constructed from such single color
criteria will not only include early-type galaxies, but may also
contain even normal early-type spiral galaxies at $z\sim 1.6 -
1.8$. The largest contamination is expected to be present in the $J-K
\geq 1.8$ sample, given the relatively small range of $J-K$ color
observed among galaxies of all morphological types. As discussed
above, only after obtaining spectroscopic and morphological
information it will be possible to produce a sample of bona fide high
redshift field elliptical galaxies.

\begin{table*}
  \caption{EROs samples selected using different color criteria}
  \label{tab:samples}
\begin{tabular}{lrrrr}
\hline \hline
    \multicolumn{5}{c}{$RJKs$ area (43 square arcmin)} \\
    $Ks^{lim}$ & N($<Ks^{lim}$) & N($J-Ks>1.8$) & N($R-Ks>5.0$) & 
    N($R-Ks>6.0$)\\ 
\hline 
    19.0 & 185 &  30 &  4 &  1 \\
    19.5 & 270 &  57 & 16 &  2 \\
    20.0 & 362 &  87 & 22 &  4 \\
    20.5 & 488 & 122 & 36 &  8 \\
    21.0 & 692 & 190 & 54 & 17 \\
\hline \hline
    \multicolumn{5}{c}{$RIJKs$ area (23.5 square arcmin)} \\
    $Ks^{lim}$ & N($<Ks^{lim}$) & N($I-Ks>4.0$) & N($I-Ks>5.0$) & 
    N($R-Ks>5.0$)\\ 
\hline 
    19.0 &  99 &   5 &  0 &  3 \\
    19.5 & 149 &  14 &  0 &  9 \\
    20.0 & 200 &  20 &  2 & 11 \\
    20.5 & 280 &  33 &  8 & 21 \\
    21.0 & 397 &  56 & 11 & 30 \\
\hline \hline
  \end{tabular}
\end{table*}

Using the multi-band observations in the 
AXAF (Chandra) Deep Field described in the previous section, we cover 
a large enough area to derive a reliable estimate of the EROs surface 
density to different limiting magnitudes, and we can also compare the 
results obtained adopting different selection criteria.  
Table~\ref{tab:samples} presents an overview of the different samples 
we have extracted from the available data using single-color criteria.

The properties of the total sample of $Ks$-selected objects agree quite
well with those of previously published samples.  Over the magnitude
interval $19.5 \leq Ks \leq 21.0$ the logarithmic slope of the objects
number counts is approximately 0.27, comparable to those derived by
Moustakas \etal (1997), by Djorgovski \etal (1995), and by Gardner,
Cowie \& Wainscoat (1993). The surface density of objects brighter than
$Ks = 20.0$ is approximately 8.5 arcmin$^{-2}$, intermediate between
those derived by Moustakas \etal (1997) and by Cowie \etal (1996), and
that reported by Djorgovski \etal (1995). We note that the most
discrepant densities with respect to the one measured here, those
reported by Moustakas \etal (1997) and by Djorgovski \etal (1995), were
obtained from observation covering an area less than one twentieth of
that used in this work. 

The exact nature of EROs is still very uncertain because they are 
only detected at relatively faint infrared magnitudes, and are 
therefore extremely faint at optical wavelengths because of their red 
colors (see, for example, Fig. 9 in Cowie \etal 1996).  We find that 
all three EROs selection criteria adopted here produce samples that 
are significantly underpopulated of bright objects, with respect to 
the total sample of $Ks$-selected objects.  The logarithmic slope of 
the number counts of $J-Ks > 1.8$, $I-Ks > 4.0$, and $R-Ks > 5.0$ 
objects are 0.34, 0.39, and 0.36 respectively, over the $19.5\leq 
Ks\leq 21.0$ magnitude interval.  There are no EROs in our samples 
brighter than $Ks = 18$.  This fact, coupled with the observation that 
the average color of the galaxy population shows a definite blueing 
trend at these faint magnitudes (resulting from a stronger star 
formation activity in $z\sim 1$ galaxies when compared to local ones), 
and that redshift surveys limited to $K = 18$ do not include any $z > 
1$ objects (Songaila \etal 1994; Cohen \etal 1999), can be considered 
as strong circumstantial evidence in favor of EROs being, for the most 
part, quiescent high redshift objects.

The infrared criterion $J-Ks > 1.8$ appears to be the least
restrictive of the criteria used to define EROs, producing a sample of
87 (190) objects brighter than $Ks$ of 20.0 (21.0) over the 43 square
arcmin covered by the $RJKs$ observations.  The corresponding surface
brightness is 2.0 (4.5) objects per square arcmin.  On the contrary,
the $R-Ks > 5.0$ criterion appears to be the most restrictive one,
producing a sample of 22 (54) objects brighter than $Ks$ of 20.0
(21.0) over the same area, that correspond to a surface density of 0.5
(1.2) objects per square arcmin.  The $I-Ks > 4.0$ criterion produces
a sample size intermediate between the previous two sample sizes.
This sample is composed of 20 (56) objects brighter than $Ks$ of 20.0
(21.0) over the 23.5 square arcmin covered by the $RIJKs$
observations.  The corresponding surface density is 0.8 (2.4) objects
per square arcmin.  The more extreme $R-Ks > 6.0$ and $I-Ks > 5.0$
selection criteria produce samples of 4 (17) and 2 (11) objects
brighter than $Ks$ of 20.0 (21.0), corresponding to surface densities
of approximately 0.1 (0.45) objects per square arcmin.

It is clear from the comparison with previously published EROs surveys
that the measurement of their average surface density is still
affected by large uncertainties, most likely due to the combined
effect of the large-scale clustering and small survey areas (see
Table~\ref{tab:surveys}).  The surface density of $R-Ks > 5.0$ objects
brighter than $Ks = 20.0$ that we have derived is smaller by a factor
of 2.5 than that reported by Cohen \etal (1999), measured over an area
approximately one third as large as the one we use in this work.  At
the same time we find a surface density of $I-Ks > 4.0$ objects
brighter than $Ks = 20.0$ more than three times higher than that
measured by Barger \etal (1999) over an area approximately three times
as large as the one we use. Assuming Poisson statistics are the only
source of uncertainty in the measurements, these differences are both
significant at approximately the $3\sigma$ level (99.36 \% and 99.75 \%
significance level, respectively). We remark that both those studies
are based on samples limited to a $K$-band magnitude of 20.0, and
therefore the corresponding densities of EROs could be affected by
subtle ``edge of the catalog'' effects.  A good agreement is found
instead with the density of $I-Ks > 4.0$ objects measured by Cowie
\etal (1996) over an area of similar extent to the one we use here.  

Taken all together these results indicate that it might be premature 
to derive any conclusion on the density evolution and on the formation 
epoch of elliptical galaxies from the currently available EROs 
surveys. For example, based on their $I-K$ selected sample, Barger 
\etal (1999) concluded that volume density of high-$z$ field 
ellipticals could be no more than 50\% of the current epoch volume 
density.  But as Table~\ref{tab:surveys} illustrates, using similar 
selection criteria, we derive an EROs surface density three times 
larger than the Barger \etal result.  Simply scaling by the implied 
Barger \etal surface-to-volume density ratio, our analogous upper 
limit on high-$z$ field ellipticals is larger than the current epoch 
value.  Only measuring the true redshift distribution of both samples 
will resolve this contradiction.

\section{A multi-color selected EROs sample}
\label{sec:ellipticals}

In the previous section, we discussed EROs samples constructed using a
single-color criterion.  Such single-color samples will obviously
contain interlopers to a greater or lesser degree for one simple
reason: one color is not enough to uniquely determine the complete SED
of any given object.  This simple fact has driven modern photometric
redshift surveys to increasing numbers of filters to more accurately
constrain the measured SED (\eg Koo 1984).  In a similar vein, we can
use the three colors available to us ($R-K$, $I-K$, and $J-K$) to
construct an EROs sample which should have a higher yield of actual
high-redshift elliptical galaxies.  We are applying a very
straightforward principle here: high-redshift (z $>$ 1) passively
evolving luminous ellipticals will tend to satisfy all three criteria
simultaneously while most sample interlopers (\eg dusty starburst
galaxies, M-type dwarf stars) will not. Therefore, our $RIJK$ sample
will better constrain the high-redshift elliptical galaxy volume
density upper limit, although it might still contain a number of
early-type spiral galaxies with redshift $z\sim 1.6 - 1.8$, as
discussed in the previous section.

\begin{table*}
  \caption{Comparison between single- and multi-color EROs selection criteria}
  \label{tab:efficiency}
\begin{tabular}{lrrrr}
\hline \hline
    \multicolumn{5}{c}{$RIJKs$ area (23.5 square arcmin)} \\
    $Ks_{lim}$ & N($J-Ks>1.8$) & N($I-Ks>4.0$) & N($R-Ks>5.0$) & N(E,$z>1$) \\ 
\hline 
    19.0 &  15 &  5 &  3 &  3 \\
    19.5 &  25 & 14 &  9 &  8 \\
    20.0 &  39 & 20 & 11 & 10 \\
    20.5 &  64 & 33 & 21 & 17 \\
    21.0 & 100 & 56 & 30 & 23 \\
\hline \hline
  \end{tabular}
\end{table*}

Table~\ref{tab:efficiency} presents the results of the application of 
the multi-color criteria to the object catalog from the area covered 
by the $RIJKs$ observations, and compares it with the results of the 
application of the single-color criteria.  We identify a sample of 10 
(23) objects brighter than $Ks$ = 20.0 (21.0) that satisfy all three 
color criteria, over an area of 23.5 square arcmin.  This corresponds 
to a surface density of 0.4 (1.0) objects per square arcmin.  This 
density, although significantly smaller than the $I-Ks > 4$ sample, is 
still approximately 1.6 times higher than that reported by Barger
\etal (1999).  Using the Barger \etal surface-to-volume
density ratio again, we could conclude that our $RIJK$ selected sample 
implies a high-redshift volume density comparable to the current 
volume density, i.e. that significant number evolution has not occurred 
since z $\sim$ 1.5.

\begin{figure}
\resizebox{\columnwidth}{!}{\includegraphics{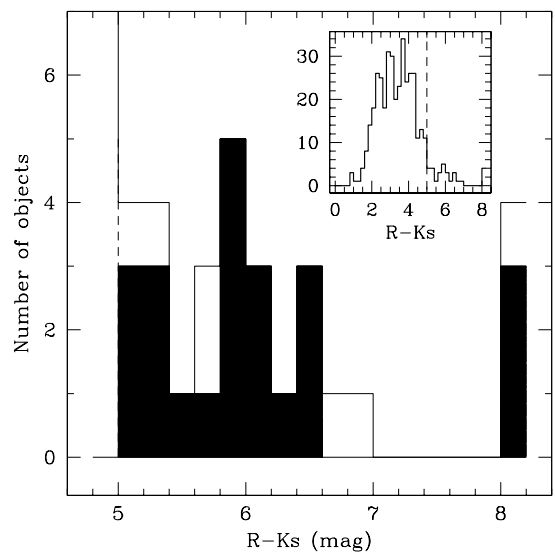}} 
\caption{The $R-Ks$ color distribution for $Ks \leq 21, R-Ks > 5$ 
objects in the area covered by the $RIJK$ data (open histogram) 
compared to the color distribution of the objects selected using the 
more restrictive $RIJK$ criteria (shaded histogram).  The latter 
objects are the most likely candidates for $z>1$ passively evolving 
elliptical galaxies.  The inset shows the color distribution for the 
entire $Ks \leq 21.0$ sample in the same area.  The vertical dashed 
line indicates the $R-Ks>5.0$ single-band EROs selection criterion.  For 
all histograms, the bin size is 0.2 mag.  Objects in the reddest bin 
represent non-detections in the $R$-band, and are given an arbitrary 
very red color purely for display purposes.}
\label{fig:rk_histo}
\end{figure}

\begin{figure}
\resizebox{\columnwidth}{!}{\includegraphics{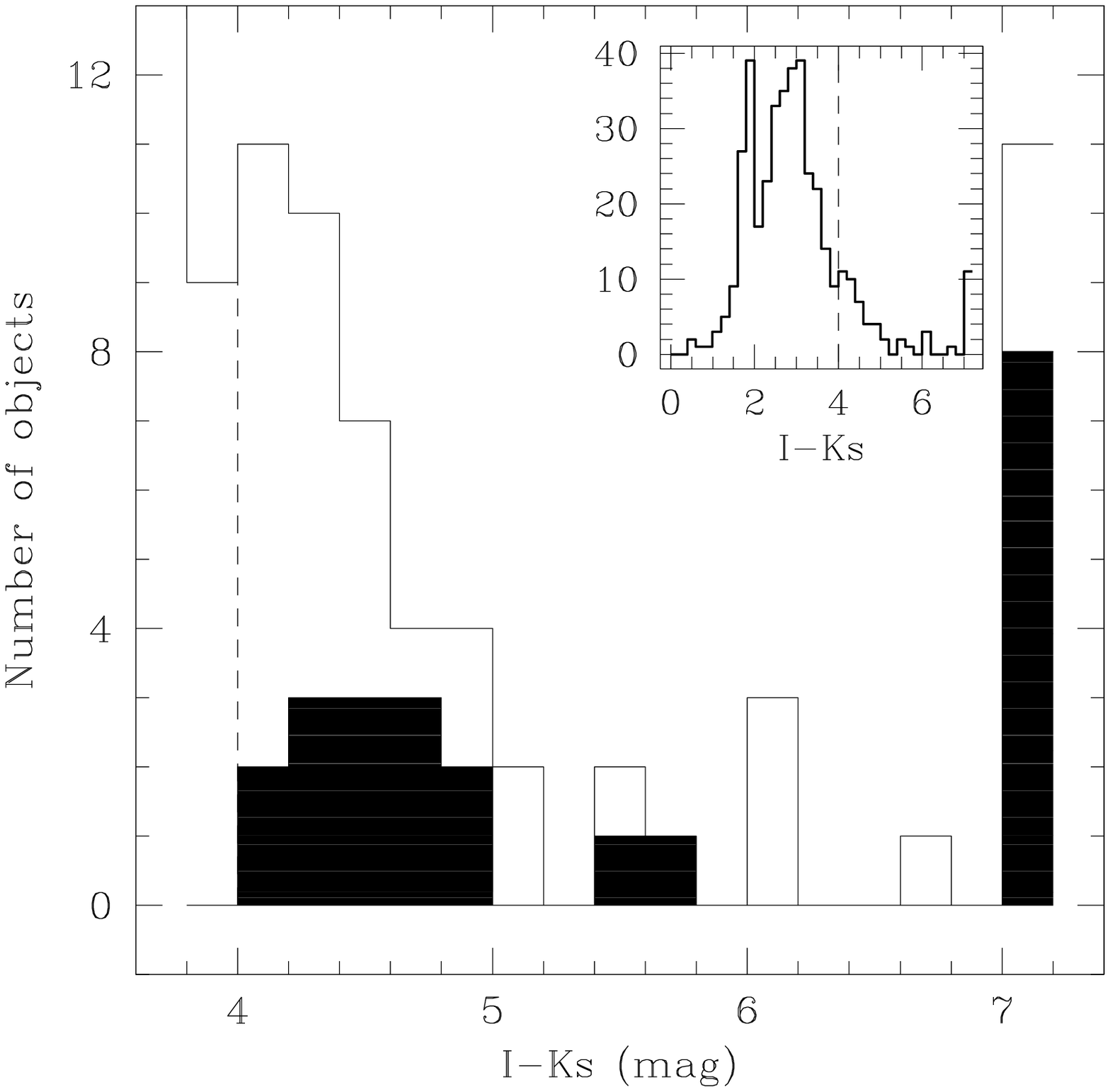}} 
\caption{The $I-Ks$ color distribution for $Ks \leq 21, I-Ks > 4$ 
objects in the area covered by the $RIJK$ data (open histogram) 
compared to the color distribution of the objects selected using the 
more restrictive $RIJK$ criteria (shaded histogram).  The latter 
objects are the most likely candidates for $z>1$ passively evolving 
elliptical galaxies.  The notation is the same as in the previous 
figure.  The inset shows the color distribution for the entire $Ks 
\leq 21.0$ sample in the same area.  The vertical dashed line 
indicates the $I-Ks>4.0$ selection criterion.  Objects in the reddest 
bin represent non-detections in the $I$-band.}
\label{fig:ik_histo}
\end{figure}

\begin{figure}
\resizebox{\columnwidth}{!}{\includegraphics{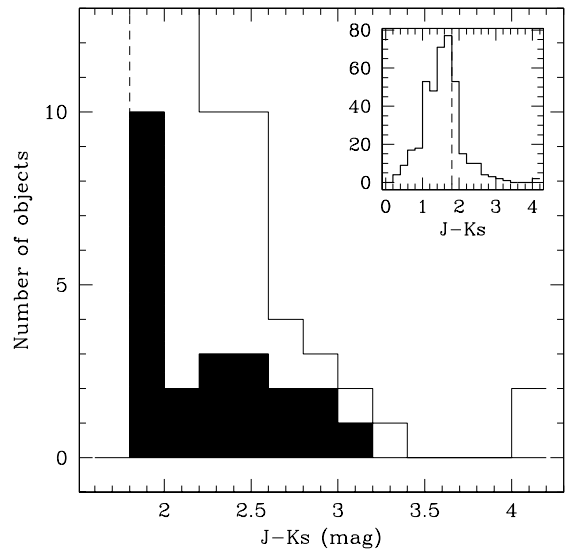}} 
\caption{The $J-Ks$ color distribution for $Ks \leq 21, J-Ks > 1.8$ 
objects in the area covered by the $RIJK$ data (open histogram) 
compared to the color distribution of the objects selected using the 
more restrictive $RIJK$ criteria (shaded histogram).  The latter 
objects are the most likely candidates for $z>1$ passively evolving 
elliptical galaxies.  The notation is the same as in the previous 
figure.  The inset shows the color distribution for the entire $Ks 
\leq 21.0$ sample in the same area.  The vertical dashed line 
indicates the $J-Ks>1.8$ selection criterion.  Objects in the reddest 
bin represent non-detections in the $J$-band.}
\label{fig:jk_histo}
\end{figure}

We can also use our multi-band selected sample to quantify the likely 
contamination fraction in the single-band selected samples discussed 
in the last section.  Consider Table~\ref{tab:efficiency} and 
Figs.~\ref{fig:rk_histo}--~\ref{fig:jk_histo}.  Recall that all 
four bands are only available for the $\sim$ 23.5 square arcmin area.  
To a limiting magnitude of $Ks = 21.0$, the 23 very likely $z > 1$ 
passively evolving elliptical galaxies are extracted from samples of 
30 $R-Ks \geq 5.0$ objects (77\% efficiency); 56 $I-Ks \geq 4.0$ 
objects (41\% efficiency) and 100 $J-Ks \geq 1.8$ objects (23\%
efficiency).  The level of contamination is therefore very low in the 
$R-Ks \geq 5.0$ sample (around 20\%), and increases dramatically when 
we move to colors based on filters with a smaller wavelength 
difference between them, reaching approximately 75\% level for the 
$J-Ks > 1.8$ sample.

Can we conclude that $R-Ks > 5.0$ is the best criterion to search for 
high-redshift ellipticals when no other information is available?  
This question will be addressed further in Papers 2 and 3, as we study 
other EROs samples and do a more extensive color plane analysis of 
possible contaminating populations using spectral synthesis models.

We close by remarking that there is no evidence for high-redshift
clusters in the 23.5 square arcmin area covered by our $RIJK$ selected
sample.  However, there appears to be one high-redshift cluster
candidate in the total $RJK$ area.  This topic is discussed more
extensively in Scodeggio \etal (2000, in preparation).

\section{Summary}
\label{sec:summary}

We have used the publicly available EIS-DEEP optical-NIR data for 
the AXAF (Chandra) Deep Field to construct samples of Extremely Red 
Objects (EROs) using various color criteria consistent with passively 
evolving elliptical galaxies at z $>$ 1.  Our conclusions can be 
summarized as follows:

\begin{enumerate}

\item{} The surface density of EROs is still far from being reliably 
determined, and there are significant differences in different 
surveys.  The EROs surface densities derived here lie between the 
values derived for other studies.  A larger, and preferably 
non-contiguous, area should be used to better constrain EROs surface 
density.

\item{} Just one color measurement is not enough to select accurately 
likely $z > 1$ elliptical galaxies.  However, if obtaining more than 
one color is impossible or undesirable for some reason, our results 
suggest that $R-K$ is the most efficient color to use.  We will study 
this issue further in our next paper.

\item{} We have used multi-band EROs selection criterion to construct a 
sample of likely $z > 1$ passively evolving elliptical galaxies.  Our 
derived surface density is consistent with little or no volume density 
of such objects since z $\sim$ 1.5.  This result is tenuous, as well 
as being based on very simple arguments, and requires spectroscopic 
followup to confirm the nature of the multi-band selected EROs 
population.

\end{enumerate}

\begin{acknowledgements}

We thank R{\"o}land Rengelink and Mario Nonino of the EIS Team for
their assistance in deriving the revised chi-square images and object
catalogues, as well as the entire EIS Team for their efforts in
obtaining and distributing the data used for this study.
 
\end{acknowledgements}


\begin{thebibliography}{}

\bibitem[]{} Andreani, P., Cimatti, A., R{\"o}ttgering, H., 
	Tilanus, R., 1999, astro-ph/9903121
\bibitem[]{} Barger, A.J., Cowie, L.L., Trentham, N., Fulton, E.,
	Hu, E.M., Songaila, A., Hall, D., 1999, AJ 117, 102
\bibitem[]{} Bertin, E., Arnouts, S., 1996, A\&AS 117, 393
\bibitem[]{} Blumenthal, G.R., Faber, S.M., Primack, J.R.,
	Rees, M.J., 1984, Nature 311, 517
\bibitem[]{} Bruzual, A.G., Charlot, S., 1993, ApJ, 405, 538 
\bibitem[]{} Cimatti, A., Andreani, P., R{\"o}ttgering, H., 
	Tilanus, R., 1998, Nature 392, 895
\bibitem[]{} Cohen, J.G., Blanford, R., Hogg, D.W., Pahre, M.A., 
	Shopbell, P.L., 1999, ApJ 512, 30
\bibitem[]{} Cowie, L.L., Songaila, A., Hu, E.M., Cohen, J.G., 1996,
	AJ 112, 839
\bibitem[]{} da Costa, L., Nonino, M., Rengelink, R., et al., 1999,
	A\&A in press (astro-ph/9812105)
\bibitem[]{} Devillard, N., 1998, Eclipse Data Analysis Software
	Package (ESO: Garching)
\bibitem[]{} Djorgovski, G.S., Soifer, B.T., Pahre, M.A., et al., 1995,
	ApJ 438, L13
\bibitem[]{} D'Odorico, S., Beletic, J.W., Amico, P., Hook, I.,
	Marconi, G., Pedichini, F., 1998, Proc. SPIE 3355, 507
\bibitem[]{} Dunlop, J., Peacock, J., Spinrad, H., Dey, A., Jimenez, R.,
 	Stern, D., Windhorst, R., 1996, Nature 381, 581
\bibitem[]{} Eggen, O.J., Lynden-Bell, D., Sandage, A., 1962, 
	ApJ 136, 748 
\bibitem[]{} Fruchter, A.S., Hook, R.N., 1999, PASP in press 
	(astro-ph/9808087)
\bibitem[]{} Gardner, J.P., Cowie, L.L., Wainscoat, R.J., 1993
	ApJ 415, L9
\bibitem[]{} Graham, J.R., Dey, A., 1996, ApJ 471, 720
\bibitem[]{} Kauffmann, G., Charlot, S., White, S.D.M., 1996, 
	MNRAS 283, L117
\bibitem[]{} Koo, D., 1985, AJ 90, 418
\bibitem[]{} Landolt, A.U., 1992, AJ 104, 340
\bibitem[]{} Larson, R.B., 1974, MNRAS 166, 585
\bibitem[]{} Lilly, S.J., Tresse, L., Hammer, F., Crampton, D.,
	LeFevre, O., 1995, ApJ 455, 108
\bibitem[]{} Moorwood, A., Cuby, J.G., Lidman, C., 1998, 
	The ESO Messenger 91, 9
\bibitem[]{} Moustakas, L.A., Davis, M., Graham, J.R., Silk, J.,
	Peterson, B., Yoshii, Y., 1997, ApJ 475, 445
\bibitem[]{} Persson, S.E., Murphy, D.C., Krzeminski, W., 
	Roth, M., Rieke, M.J., 1998, AJ 116, 2475
\bibitem[]{} Rengelink, R., Nonino, M., da Costa, L., et al., 1999, 
	A\&A, in press (astro-ph/9812190)
\bibitem[]{} Schade, D., Lilly, S.J., Crampton, D., et al., 1999, ApJ
	525, 31
\bibitem[]{} Songila, A., Cowie, L.L., Hu, E.M., Gardner, J.P., 1994, 
    ApJS 94, 461
\bibitem[]{} Spinrad, H., Dey, A., Stern, D., Dunlop, J., Peacock, J., 
    Jimenez, R., Windhorst, R., 1997, ApJ 484, 581
\bibitem[]{} Szalay, A.S., Connoly, A.J., Szokoly, G.P., 1999, 
	AJ 117, 68
\bibitem[]{} Thompson, D., Beckwith, S.V.W., Fockenbrock, R., et
	al., 1999, ApJ 523, 100
\bibitem[]{} Tinsley, B.M., Gunn, J.E., 1976, ApJ 203, 52
\bibitem[]{} Toomre, A., 1977, in ``The Evolution of Galaxies and
	Stellar Populations'', ed. Tinsley, B.M., Larson, R.B. 
	(Yale Univ. Obs.: New Haven) p. 401
\bibitem[]{} White, S.D.M., Rees, M.J., 1978, MNRAS 183, 341

\end{thebibliography}
\end{document}